\documentclass[a4paper,11pt]{article}
\usepackage{cite}
\usepackage[all]{xy}
\usepackage{amsmath,amsfonts,amssymb,amsthm,epsfig,amscd,comment,latexsym,psfrag}
\usepackage{CJK}
\usepackage{mathrsfs}
\usepackage{nicefrac,xspace,tikz}
\usepackage{arydshln}
\usepackage{stmaryrd}
\usepackage{graphicx,bm}
\usepackage{amscd}
\numberwithin{equation}{section}
\usetikzlibrary{arrows}

\usepackage{graphicx,amssymb,amsmath,bm,latexsym}

\allowdisplaybreaks

\begin{document}
\begin{titlepage}
\begin{center}
{\Large\bf Correlators in the Gaussian and chiral supereigenvalue models in the Neveu-Schwarz sector}\vskip .2in
{\large Rui Wang$^{a,}$\footnote{wrui@amss.ac.cn},
 Shi-Kun Wang$^{a,}$\footnote{wsk@amss.ac.cn},
  Ke Wu$^{b,}$\footnote{wuke@cnu.edu.cn},
Wei-Zhong Zhao$^{b,}$\footnote{Corresponding author: zhaowz@cnu.edu.cn}} \vskip .2in
$^a${\em Institute of Applied Mathematics, Academy of Mathematics and Systems Science,
Chinese Academy of Sciences, Beijing 100190, China}\\
$^b${\em School of Mathematical Sciences, Capital Normal University,
Beijing 100048, China} \\

\begin{abstract}
We analyze the Gaussian and chiral supereigenvalue models in the Neveu-Schwarz sector.
We show that their partition functions can be expressed as the infinite
sums of the homogeneous operators acting on the elementary functions.
In spite of the fact that the usual $W$-representations of these matrix models
can not be provided here, we can still derive the compact expressions of the correlators
in these two supereigenvalue models. Furthermore, the non-Gaussian (chiral) cases are also discussed.
\end{abstract}

\end{center}


\end{titlepage}

\section{Introduction}

The supereigenvalue models have attracted considerable attention.
They can be regarded as supersymmetric generalizations of matrix models
\cite{Zadra}-\cite{Osuga}.
Many of the important features of matrix models,
such as the Virasoro constraints, the loop equations, the genus expansions
and the moment descriptions have their supersymmetric counterparts in the supereigenvalue models.
Two types of supereigenvalue models in the Ramond sector, i.e., Ramond-NS and Ramond-R supereigenvalue models, were
derived in Ref. \cite{Ciosmak} by means of the conformal field theory formalism.
They obey the super Virasoro constraints,
where the bosonic and fermionic operators yield the Ramond version of the super-Virasoro
algebra.  Moreover, the corresponding Ramond super quantum curves were analyzed
and the underlying Ramond super spectral curves were identified.
It is well-known that the $W$-representations can be used to realize partition functions of various matrix models,
such as the Gaussian Hermitian and complex matrix models  \cite{Shakirov2009}-\cite{Itoyama2020},
the Kontsevich matrix model \cite{Alexandrov2011}
and the generalized Brezin-Gross-Witten model \cite{Alexandrov2016}.
Namely, by acting on elementary functions with exponents of the given $W$-operators,
we can give the corresponding partition functions of the matrix models.
Recently it was proved that Ramond-NS
supereigenvalue model can be obtained in terms of the $W$-representation \cite{Chen}
\begin{eqnarray}\label{pse}
Z&=&(\prod_{i=1}^N \int_{0}^{+\infty} dz_i \int d\theta_i) \Delta_R(z,\theta)^\beta e^{-\frac{\sqrt{\beta}}{\hbar}\sum_{i=1}^{N}
(\sum_{k=0}^{\infty}(t_kz_i^k+\xi_kz_i^{k-\frac{1}{2}}\theta_i)+z_i)}\nonumber\\
&=&e^{-{W}} e^{-\frac{\sqrt{\beta}}{\hbar}Nt_0},
\end{eqnarray}
where $N$ is even, $z_i$ are bosonic variables and $\theta_i$
are Grassmann variables,
$t_k$ and $\xi_{k}$ are bosonic and fermionic coupling constants, respectively,
$\Delta_R(z,\theta)$ is the Vandermonde-like determinant,
$\Delta_R(z,\theta)=\prod_{1\leq i<j\leq N}(z_i-z_j-\frac{1}{2}(z_i+z_j)\frac{\theta_i\theta_j}{\sqrt{z_iz_j}}).$
Due to the $W$-representation, the compact expressions of correlators in the supereigenvalue model (\ref{pse})
can be derived. The final result shows that the correlators are determined by the certain coefficients
in the power of the operator $W$.

The supereigenvalue model in the Neveu-Schwarz sector describes the coupling between ($2,4m$) superconformal models and world-sheet supergravity \cite{Zadra}.
The $\beta$-deformed case of this supereigenvalue model is given by \cite{Sulkowski}
\begin{eqnarray}\label{bspse}
 \tilde Z&=&(\prod_{i=1}^N \int_{-\infty}^{+\infty} dz_i \int d\theta_i) \Delta_{NS}(z,\theta)^\beta e^{-\frac{\sqrt{\beta}}{\hbar}\sum_{i=1}^{N}V_{NS}(z_i,\theta_i)},
\end{eqnarray}
where $N$ is even,
$\Delta_{NS}(z,\theta)=\prod_{1\leq i<j\leq N}(z_i-z_j-\theta_i\theta_j),$
$V_{NS}(z,\theta)=\sum_{k=0}^{\infty}t_kz^k
+\sum_{k=0}^{\infty}\xi_{k+\frac{1}{2}}z^k
\theta,$
and $\xi_{k+\frac{1}{2}}$ are fermionic coupling constants.
There are the super Virasoro constraints for the partition function (\ref{bspse}),
where the bosonic and fermionic operators yield the Neveu-Schwarz version of the super-Virasoro
algebra.
When the bosonic variables $z_i$ in (\ref{bspse}) are integrated from $0$ to $+\infty$,
it gives the chiral supereigenvalue model,
which can be seen as the supersymmetric generalization
of the complex one-matrix model.
The closed expressions for all planar multi-superloop correlation functions were derived for $\beta=1$ chiral supereigenvalue model. It was found that
the higher genus contributions to the free energy and the correlators
can be calculated by an iterative scheme \cite{Akemann}.

It is known that the compact expressions of the correlators in the supereigenvalue model
(\ref{pse}) in the Ramond sector have been presented.
However, for the cases of the supereigenvalue model in the Neveu-Schwarz sector,
it still remains to be seen whether there are the similar results.
In this paper, we analyze the (non-)Gaussian supereigenvalue model in the Neveu-Schwarz sector and
give the correlators in these matrix models. Moreover, the cases of the chiral supereigenvalue model
will be also investigated.

This paper is organized as follows. In section 2, we focus on the Gaussian supereigenvalue model
in the Neveu-Schwarz sector and show that its partition function can be expressed as the infinite
sums of the homogeneous operators acting on the elementary functions.
Moreover, we derive the compact expressions of the correlators.
In section 3, we consider the non-Gaussian supereigenvalue model in the Neveu-Schwarz sector.
In section 4, we consider the chiral supereigenvalue model and present
the compact expressions of the correlators.
The non-Gaussian chiral case is analyzed in section 5.
We end this paper with the conclusions in section 6.

\section{Gaussian supereigenvalue model in the Neveu-Schwarz sector}

The Gaussian Hermitian matrix model is one of the most studied and best understood matrix models.
Its partition function can be generated by the $W$-representation \cite{Shakirov2009}.
The correlators in this matrix model have been well discussed \cite{Harer}-\cite{Kang} and
the compact expressions of the correlators have been presented
\cite{Itoyama17}-\cite{Kang}.
For the non-Gaussian Hermitian matrix model,  the correlators can be evaluated by
the recursive formulas derived from the Virasoro constraints and the additional constraints.
However, it is hard to give the compact expressions of the correlators
\cite{Alexandrov04}-\cite{2007.10354}.

In this section, we focus on the Gaussian supereigenvalue model which is obtained by taking the shift
$t_2\rightarrow t_2+\frac{1}{2}$ in the potential of the partition function
(\ref{bspse})
\begin{eqnarray}\label{gnspf}
 Z_G&=&\frac{1}{\Lambda_G}
 (\prod_{i=1}^N \int_{-\infty}^{+\infty} dz_i \int d\theta_i)\Delta_{NS}(z,\theta)^\beta e^{-\frac{\sqrt{\beta}}{\hbar}\sum_{i=1}^{N}
 (V_{NS}(z_i,\theta_i)+\frac{1}{2}z^2_i)},
\end{eqnarray}
where the normalization factor $\Lambda_G$ is given by
\begin{eqnarray}
\Lambda_G=(\prod_{i=1}^N \int_{-\infty}^{+\infty} dz_i \int d\theta_i) \Delta_{NS}(z,\theta)^\beta e^{-\frac{\sqrt{\beta}}{2\hbar}
\sum_{i=1}^{N}z^2_i}.
\end{eqnarray}

From the invariance of the integral (\ref{gnspf}) under the changes of integration
variables $(z_i\rightarrow z_i+\epsilon z_i^{n+1}, \quad
\theta_i\rightarrow \theta_i+\frac{1}{2}\epsilon (n+1)\theta_iz_i^n)$ and
$(z_i\rightarrow z_i+ z_i^{n+1}\theta_i\delta, \quad \theta_i\rightarrow
\theta_i+ z_i^{n+1}\delta)$,  where $\epsilon$ and $\delta$ are the
infinitesimal bosonic and fermionic constants, respectively, we can give
the super Virasoro constraints
\begin{eqnarray}\label{bssvcpse}
\ell_n Z_G=0,\quad\quad  g_{n+\frac{1}{2}}Z_G=0, \quad\quad
n\geq-1,
\end{eqnarray}
where
\begin{eqnarray}
\ell_n&=&\frac{\partial}{\partial t_{n+2}}+\sum_{k=1}^\infty kt_k\frac{\partial}{\partial t_{n+k}}+\frac{\hbar^2}{2}\sum_{k=0}^n\frac{\partial}{\partial t_k}\frac{\partial}{\partial t_{n-k}}+\sum_{k=0}^\infty(k+\frac{n+1}{2})\xi_{k+\frac{1}{2}}\frac{\partial}{\partial \xi_{k+n+\frac{1}{2}}} \nonumber\\
 &&+\frac{\hbar^2}{2}\sum_{k=1}^n k\frac{\partial}{\partial \xi_{n-k+\frac{1}{2}}}\frac{\partial}{\partial \xi_{k-\frac{1}{2}}}-\frac{\hbar}{2\sqrt{\beta}}(1-\beta)(n+1)\frac{\partial}{\partial t_n},\nonumber\\
 g_{n+\frac{1}{2}}&=&\frac{\partial}{\partial \xi_{n+\frac{5}{2}}}+
 \sum_{k=1}^\infty kt_k\frac{\partial}{\partial \xi_{n+k+\frac{1}{2}}}+\sum_{k=0}^\infty\xi_{k+\frac{1}{2}}\frac{\partial}{\partial t_{k+n+1}}+\hbar^2\sum_{k=0}^n\frac{\partial}{\partial \xi_{k+\frac{1}{2}}}\frac{\partial}{\partial t_{n-k}}\nonumber\\
 &&-\frac{\hbar}{\sqrt{\beta}}(1-\beta)(n+1)\frac{\partial}{\partial \xi_{n+\frac{1}{2}}}.
\end{eqnarray}
The constraint operators obey the Neveu-Schwarz version of the super-Virasoro algebra
\begin{eqnarray}
&&[ \ell_m, \ell_n]=(m-n) \ell_{m+n},\nonumber\\
&&[ \ell_m, g_{n+\frac{1}{2}}]=(\frac{m-1}{2}-n) g_{m+n+\frac{1}{2}},\nonumber\\
&&\{g_{m+\frac{1}{2}}, g_{n+\frac{1}{2}}\}=2\ell_{m+n+1}.
\end{eqnarray}

Let us now take the change of variables given by
\begin{eqnarray}\label{btran1}
&&z_i\rightarrow z_i+\epsilon\sum_{n=1}^\infty nt_{n}z_i^{n-1}, \quad \theta_i\rightarrow \theta_i+\frac{1}{2}\epsilon\sum_{n=2}^\infty n(n-1)t_{n}z_i^{n-2}\theta_i.
\end{eqnarray}
By requiring that the partition function is invariant under the infinitesimal
transformations (\ref{btran1}), it leads to the constraint
\begin{equation}\label{loop1}
(\tilde D_1+\tilde W_1)Z_G=0,
\end{equation}
where
\begin{eqnarray}\label{dw1}
\tilde{D}_1&=&\sum_{k=1}^\infty kt_k\frac{\partial}{\partial t_k},
\nonumber\\
\tilde{W}_1&=&\sum_{n,k=1}^{\infty}nkt_nt_k\frac{\partial}{\partial t_{n+k-2}}+\sum_{n=1}^{\infty}\sum_{k=0}^{\infty}n(k+\frac{n-1}{2})t_n
\xi_{k+\frac{1}{2}}
\frac{\partial}{\partial \xi_{n+k-\frac{3}{2}}} \nonumber\\
&&+\frac{\hbar^2}{2}\sum_{n=2}^\infty\sum_{k=0}^{n-2}nt_n
\frac{\partial}{\partial t_{k}}\frac{\partial}{\partial t_{n-k-2}}
+\frac{\hbar^2}{2}\sum_{n=3}^\infty\sum_{k=1}^{n-2}nkt_n
\frac{\partial}{\partial \xi_{n-k-\frac{3}{2}}}\frac{\partial}{\partial \xi_{k-\frac{1}{2}}}  \nonumber\\
&&-\frac{\hbar}{2\sqrt{\beta}}(1-\beta)\sum_{n=2}^\infty n(n-1)t_n
\frac{\partial}{\partial t_{n-2}}.
\end{eqnarray}

Similarly, by requiring the invariance of the partition function (\ref{gnspf})  under
\begin{eqnarray}\label{btran2}
&&z_i\rightarrow z_i+\epsilon\sum_{n=1}^\infty (n+\frac{1}{2})
\xi_{n+\frac{1}{2}}z_i^{n-1}\theta_i, \quad \theta_i\rightarrow \theta_i-\epsilon\sum_{n=1}^\infty (n+\frac{1}{2})\xi_{n+\frac{1}{2}}z_i^{n-1},
\end{eqnarray}
it gives another constraint
\begin{equation}\label{loop2}
(\tilde D_2+\tilde W_2)Z_G=0,
\end{equation}
where
\begin{eqnarray}\label{dw2}
\tilde{D}_2&=&\sum_{k=1}^\infty (k+\frac{1}{2})\xi_{k+\frac{1}{2}}\frac{\partial}{\partial \xi_{k+\frac{1}{2}}},
\nonumber\\
\tilde{W}_2&=&\sum_{n,k=1}^{\infty}n(k+\frac{1}{2})t_n\xi_{k+\frac{1}{2}}
\frac{\partial}{\partial \xi_{n+k-\frac{3}{2}}}
-\frac{\hbar}{\sqrt{\beta}}(1-\beta)\sum_{n=2}^\infty (n-1)(n+\frac{1}{2})\xi_{n+\frac{1}{2}}\frac{\partial}{\partial \xi_{n-\frac{3}{2}}} \nonumber\\
&&+\hbar^2\sum_{n=2}^\infty
\sum_{k=0}^{n-2}(n+\frac{1}{2})\xi_{n+\frac{1}{2}}
\frac{\partial}{\partial \xi_{k+\frac{1}{2}}}\frac{\partial}{\partial t_{n-k-2}} +\sum_{n=1}^{\infty}\sum_{k=0}^{\infty}(n+\frac{1}{2})\xi_{n+\frac{1}{2}}
\xi_{k+\frac{1}{2}}\frac{\partial}{\partial t_{n+k-1}} .\nonumber\\
\end{eqnarray}
Combining (\ref{loop1}) and (\ref{loop2}), we have
\begin{equation}\label{ccons}
(\tilde D+\tilde W)Z_G=0,
\end{equation}
where $\tilde D=\tilde D_1+\tilde D_2$, $\tilde W=\tilde W_1+\tilde W_2$.

The partition function (\ref{gnspf}) can be formally expanded as
\begin{eqnarray}\label{expan}
Z_G&=&e^{-\frac{\sqrt{\beta}}{\hbar}Nt_{0}}+\sum_{s=1}^{\infty}Z_G^{(s)}\nonumber\\
&=&e^{-\frac{\sqrt{\beta}}{\hbar}Nt_{0}}\Big[1-\frac{\sqrt{\beta}}{\hbar}
\tilde C_{k_1}t_{k_1}
+\frac{1}{2!}
(\frac{\sqrt{\beta}}{\hbar})^2 \tilde C_{k_1, k_2}t_{k_1}t_{k_2}
-\frac{1}{2!}(\frac{\sqrt{\beta}}{\hbar})^2
\tilde C^{s_1+\frac{1}{2}, s_2+\frac{1}{2}}\xi_{s_1+\frac{1}{2}}\xi_{s_2+\frac{1}{2}}\nonumber\\
&&-\frac{1}{3!}(\frac{\sqrt{\beta}}{\hbar})^3 \tilde C_{k_1, k_2, k_3}
t_{k_1}t_{k_2}t_{k_3}+\frac{1}{2!}(\frac{\sqrt{\beta}}{\hbar})^3
\tilde C_{k_1}^{s_1+\frac{1}{2}, s_2+\frac{1}{2}}t_{k_1}\xi_{s_1+\frac{1}{2}}\xi_{s_2+\frac{1}{2}}+\cdots\Big],
\end{eqnarray}
where
\begin{eqnarray}\label{deg}
&&Z_G^{(s)}=e^{-\frac{\sqrt{\beta}}{\hbar}Nt_0}\Big[\sum_{n=0}^{\infty}
\sum_{\substack{m=0\\m\ \text{is even}}}^{\infty}
\sum_{\substack{k_1+\cdots +k_n+\\s_1+\cdots+s_m+\frac{m}{2}=s\\k_1,\cdots,k_n\geq 1\\s_1,\cdots,s_m\geq 0}}
\frac{(-1)^{\frac{m(m+1)}{2}}(-\frac{\sqrt{\beta}}{\hbar})^{n+m}}{n!m!}
\tilde C_{k_1, \cdots, k_n}^{s_1+\frac{1}{2}, \cdots, s_m+\frac{1}{2}}\nonumber\\
&&\cdot t_{k_1}\cdots t_{k_n}\xi_{s_1+\frac{1}{2}}\cdots\xi_{s_m+\frac{1}{2}}\Big],
\end{eqnarray}
and the coefficients $\tilde C_{k_1, \cdots, k_n}^{s_1+\frac{1}{2}, \cdots, s_m+\frac{1}{2}}$ are the correlators defined by
\begin{eqnarray}\label{decor}
\tilde C_{k_1, \cdots, k_n}^{s_1+\frac{1}{2}, \cdots, s_m+\frac{1}{2}}&=&\frac{1}{\Lambda_G}(\prod_{i=1}^N \int_{-\infty}^{+\infty} dz_i \int d\theta_i)
\sum_{\substack{i_1,\cdots, i_n=1\\j_1,\cdots, j_m=1}}^N
z_{i_1}^{k_1}\cdots z_{i_n}^{k_n}
z_{j_1}^{s_1}\theta_{j_1}\cdots
z_{j_m}^{s_m}\theta_{j_m}\nonumber\\
&&\cdot \Delta_{NS}(z,\theta)^\beta e^{-\frac{\sqrt{\beta}}{2\hbar}\sum_{i=1}^{N}z^2_i}.
\end{eqnarray}

Let $V$ be the infinite dimensional vector space on $\mathbb{C}[[t_0]]$
 generated by the basis
\begin{equation}\label{vbasis}
\{t_{k_1}\cdots t_{k_n}\xi_{s_1+\frac{1}{2}}\cdots\xi_{s_m+\frac{1}{2}}|
k_1,\cdots,k_n\geq 1, s_1,\cdots,s_m\geq 0, n\in \mathbb{N}, m\in 2\mathbb{N}\}.
\end{equation}
Thus the partition function $Z_G$ can be seen as a vector in $V$, the operators $\tilde D$ and $\tilde W$ are differential operators on $V$.
Under the natural gradation $deg(t_k)=k$, $deg(\frac{\partial}{\partial t_k})=-k$, $deg(\xi_{k+\frac{1}{2}})=k+\frac{1}{2}$, and $deg(\frac{\partial}{\partial \xi_{k+\frac{1}{2}}})=-k-\frac{1}{2}$,
$k=0,1,\cdots, \infty$,
we have $deg(\tilde D)=0$ and $deg(\tilde W)=2$. In addition, the kernel of $\tilde D$ denoted as $Ker(\tilde D)=\{v\in V|\tilde Dv=0\}$, is one dimensional.
It is not difficult to see that $Ker(\tilde D+\tilde W)$ is still one dimensional. Therefore, the partition function $Z_G$ is uniquely determined by the constraint (\ref{ccons}).
As a consequence, the correlators (\ref{decor}) can be totally derived from (\ref{ccons}).

Let us collect the coefficients of $t_1^{l}$ and $t_2^{l}$
in (\ref{ccons}) and set to zero, respectively, we obtain
\begin{equation}
\tilde C_{1}=0,\quad \tilde C_{1,1}=\frac{\hbar}{\sqrt{\beta}}N,
\quad \tilde C_{2}=\frac{\hbar}{2\sqrt{\beta}}N\tilde N,
\end{equation}
and the recursive relations
\begin{eqnarray}
&&\tilde C_{\underbrace{1, \cdots, 1}_{l}}=\frac{\hbar}{\sqrt{\beta}}
N(l-1)\tilde C_{\underbrace{1, \cdots, 1}_{l-2}},\nonumber\\
&&\tilde C_{\underbrace{2, \cdots, 2}_{l}}=\frac{\hbar}{\sqrt{\beta}}
[\frac{N\tilde N}{2}+2(l-1)]\tilde C_{\underbrace{2, \cdots, 2}_{l-1}},
\end{eqnarray}
where $\tilde{N}=\beta N+1-\beta$.

Then it is easy to obtain
\begin{eqnarray}\label{gc111}
\tilde C_{\underbrace{1, \cdots, 1}_{l}}
=\left\{
\begin{array}{ll}
(\frac{\hbar}{\sqrt{\beta}}N)^{\frac{l}{2}}(l-1)!!, & \hbox{for $l$  even;} \\
\\
0, & \hbox{for $l$ odd,}
\end{array}
\right.
\end{eqnarray}
and
\begin{eqnarray}\label{gc22}
\tilde C_{\underbrace{2, \cdots, 2}_{l}}=(\frac{\hbar}{\sqrt{\beta}})^{l}
\prod_{j=0}^{l-1}(\frac{N\tilde{N}}{2}+2j).
\end{eqnarray}

Similarly, by collecting the coefficients of $t_1^lt_2$,
$t_1^l\xi_{\frac{1}{2}}\xi_{\frac{3}{2}}$, $t_2^lt_1^2$ and
$t_2^l\xi_{\frac{1}{2}}\xi_{\frac{3}{2}}$ in (\ref{ccons}) and setting to zero, respectively,
we may also obtain the corresponding recursive relations
and derive the exact correlators
\begin{eqnarray}\label{gcexs1}
\tilde C_{2,\underbrace{1, \cdots, 1}_{l}}
&=&\left\{
     \begin{array}{ll}
       (\frac{\hbar}{\sqrt{\beta}})^{\frac{l}{2}+1}
(l-1)!!(l+\frac{N\tilde{N}}{2})N^{\frac{l}{2}}, & \hbox{for $l$ even;} \\
\\
     0, & \hbox{for $l$ odd,}
     \end{array}
   \right.
\nonumber\\
\tilde C^{\frac{1}{2},\frac{3}{2}}_{\underbrace{1,\cdots, 1}_{l}}&=&
\left\{
  \begin{array}{ll}
    (\frac{\hbar}{\sqrt{\beta}}N)^{\frac{l}{2}+1}(l-1)!!, & \hbox{for $l$ even;} \\
\\
    0, & \hbox{for $l$ odd,}
  \end{array}
\right.\nonumber\\
\tilde C_{1,1,\underbrace{2, \cdots, 2}_{l}}&=&
\tilde C^{\frac{1}{2},\frac{3}{2}}_{\underbrace{2,\cdots, 2}_{l}}=
(\frac{\hbar}{\sqrt{\beta}})^{l+1}N
\prod_{j=1}^l(\frac{N\tilde{N}}{2}+2j).
\end{eqnarray}

It should be mentioned that we can also calculate the correlators
(\ref{gc111})-(\ref{gcexs1})
from the super Virasoro constraints (\ref{bssvcpse}). However, due to the complicated form of the
recursive formulas for correlators from (\ref{bssvcpse}),
it is hard to give the general expressions of the correlators.
In order to achieve more results,
let us further analyze the constraint (\ref{ccons}).

Since $\tilde D e^{-\frac{\sqrt{\beta}}{\hbar}Nt_{0}}=0$, the constraint (\ref{ccons}) can be rewritten as
\begin{equation}\label{cconseq}
(\tilde D+\tilde W)\sum_{s=1}^{\infty}Z_G^{(s)}
=-\tilde We^{-\frac{\sqrt{\beta}}{\hbar}Nt_{0}}.
\end{equation}
It is noted that the function $-\tilde W e^{-\frac{\sqrt{\beta}}{\hbar}Nt_{0}}$ on the right hand side of (\ref{cconseq}) has degree $2$. Hence the operators $\tilde D$ and $\tilde D+\tilde W$ are invertible on $-\tilde W e^{-\frac{\sqrt{\beta}}{\hbar}Nt_{0}}$.

From (\ref{cconseq}), we have
\begin{eqnarray}
\sum_{s=1}^{\infty}Z_G^{(s)}
&=&-(\tilde D+\tilde W)^{-1}\tilde We^{-\frac{\sqrt{\beta}}{\hbar}Nt_{0}}\nonumber\\
&=&\sum_{k=1}^{\infty}(-\tilde D^{-1}\tilde W)^{k}
e^{-\frac{\sqrt{\beta}}{\hbar}Nt_{0}}.
\end{eqnarray}
Thus the partition function (\ref{gnspf}) can be expressed as
\begin{equation}\label{operrepn}
Z_G
=\sum_{k=0}^{\infty}(-\tilde D^{-1}\tilde W)^{k} e^{-\frac{\sqrt{\beta}}{\hbar}Nt_{0}}.
\end{equation}

We denote an operator $\mathcal{O}$ on $V$ the degree operator if
$\mathcal{O}f=deg(f)f$ for any homogeneous function $f\in V$.
Note that $\tilde W$ is a homogeneous operator with degree $2$ in (\ref{operrepn}).
It is not difficult to show that if $\tilde D$ is a degree operator,
the partition function (\ref{operrepn}) can be generated by the $W$-representation.
Unfortunately, it is clear that $\tilde D$ in (\ref{operrepn})
is not the degree operator.
It causes the usual $W$-representation of the matrix model (\ref{operrepn}) to fail here.
In other words, the partition function (\ref{gnspf})
can not be obtained by acting on elementary functions with exponents of the operator $\tilde W$.
In spite of this negative result, by evaluating the action of the homogeneous operator $(-\tilde D^{-1}\tilde W)^{k}$ on the function $e^{-\frac{\sqrt{\beta}}{\hbar}Nt_{0}}$,
we can still derive the compact expressions of the correlators
from the representation (\ref{operrepn}).
In what follows let us continue to discuss the correlators.

Since $\tilde D^{-1}\tilde W$ is an operator with degree $2$, we can see from  (\ref{operrepn})
that $Z_G^{(s)}=0$, when $s$ is odd. It leads to
$\tilde C_{k_1,\cdots, k_n}^{s_1+\frac{1}{2},\cdots, s_m+\frac{1}{2}}=0$,
when $\sum_{\mu=1}^{n}k_{\mu}+\sum_{\nu=1}^ms_{\nu}+\frac{m}{2}=s$ is odd.
In order to give the general expressions of the correlators for the $s$ even case,
we need to write out the action $(-\tilde D^{-1}\tilde W)^ke^{-\frac{\sqrt{\beta}}{\hbar}Nt_{0}}$ explicitly.

The operator $\tilde D^{-1}$ can be expressed as
\begin{eqnarray}\label{opered}
\tilde D^{-1}&=&(D-\frac{1}{2}\xi_{\frac{1}{2}}\frac{\partial }{\partial \xi_{\frac{1}{2}}})^{-1}\nonumber\\
&=&D^{-1}+D^{-1}(2D-1)^{-1}\xi_{\frac{1}{2}}\frac{\partial }{\partial \xi_{\frac{1}{2}}},
\end{eqnarray}
where $D$ is the degree operator
\begin{equation}
D=\sum_{k=1}^\infty kt_k\frac{\partial}{\partial t_k}+\sum_{k=0}^\infty (k+\frac{1}{2})\xi_{k+\frac{1}{2}}\frac{\partial}{\partial \xi_{k+\frac{1}{2}}},
\end{equation}
and the relations
$[D, \xi_{\frac{1}{2}}\frac{\partial }{\partial \xi_{\frac{1}{2}}}]=0$ and $(\xi_{\frac{1}{2}}\frac{\partial }{\partial \xi_{\frac{1}{2}}})^2=\xi_{\frac{1}{2}}\frac{\partial }{\partial \xi_{\frac{1}{2}}}$
are used to give (\ref{opered}).

Thus the operators in (\ref{operrepn}) now take the following form:
\begin{eqnarray}\label{DWG}
(\tilde D^{-1}\tilde W)^{k}
&=&(D^{-1}\tilde W+D^{-1}(2D-1)^{-1}\mathbb{\tilde W})^k\nonumber\\
&=&\sum_{r=1}^{\lfloor \frac{k+1}{2} \rfloor}\sum_{\substack{l_1+\cdots+l_{2r-1}=k\\l_1, \cdots, l_{2r-1}\geq 1}}
(T_1+T_2)
+\sum_{r=1}^{\lfloor \frac{k}{2} \rfloor}\sum_{\substack{l_1+\cdots+l_{2r}=k\\
l_1,\cdots,l_{2r}\geq 1}}(T_3+T_4),
\end{eqnarray}
where $\lfloor k \rfloor=Max\{m\in \mathbb{Z}|m\leq k\}$ is the floor function, $\mathbb{\tilde W}=\xi_{\frac{1}{2}}\frac{\partial }{\partial \xi_{\frac{1}{2}}}\tilde W$, and $T_i$, $i=1,\cdots,4$ are
\begin{eqnarray}
T_1&=&(D^{-1}\tilde W)^{l_{2r-1}}(D^{-1}(2D-1)^{-1}\mathbb{\tilde W})^{l_{2r-2}}
\cdots(D^{-1}(2D-1)^{-1}\mathbb{\tilde W})^{l_2}(D^{-1}\tilde W)^{l_1},\nonumber\\
T_2&=&(D^{-1}(2D-1)^{-1}\mathbb{\tilde W})^{l_{2r-1}}(D^{-1}\tilde W)^{l_{2r-2}}
\cdots(D^{-1}\tilde W)^{l_2}(D^{-1}(2D-1)^{-1}\mathbb{\tilde W})^{l_1},\nonumber\\
T_3&=&(D^{-1}(2D-1)^{-1}\mathbb{\tilde W})^{l_{2r}}(D^{-1}\tilde W)^{l_{2r-1}}
\cdots(D^{-1}(2D-1)^{-1}\mathbb{\tilde W})^{l_2}(D^{-1}\tilde W)^{l_1},\nonumber\\
T_4&=&(D^{-1}\tilde W)^{l_{2r}}(D^{-1}(2D-1)^{-1}\mathbb{\tilde W})^{l_{2r-1}}
\cdots(D^{-1}\tilde W)^{l_2}(D^{-1}(2D-1)^{-1}\mathbb{\tilde W})^{l_1}.
\end{eqnarray}

Using the fact that for the homogeneous function $f\in V$, $D^{-1}f=deg(f)^{-1}f$,
the action of the operator $T_1$ on the function $e^{-\frac{\sqrt{\beta}}{\hbar}Nt_{0}}$ gives
\begin{eqnarray}\label{T1act}
T_1e^{-\frac{\sqrt{\beta}}{\hbar}Nt_{0}}&=&
\frac{1}{k!2^k}\frac{1}{\prod_{i=l_1+1}^k(4i-1)}
\prod_{i=1}^{r-1}\prod_{j=1}^{l_{2i+1}}[4(l_1+\cdots+l_{2i}+j)-1]\\
&&\cdot \tilde W^{l_{2r-1}}\mathbb{\tilde W}^{l_{2r-2}}
\cdots\mathbb{\tilde W}^{l_2}\tilde W^{l_1}e^{-\frac{\sqrt{\beta}}{\hbar}Nt_{0}}\nonumber\\
&=&\frac{1}{k!2^k}\frac{1}{\prod_{i=l_1+1}^k(4i-1)}
\prod_{i=1}^{r-1}\prod_{j=1}^{l_{2i+1}}[4(l_1+\cdots+l_{2i}+j)-1]
e^{-\frac{\sqrt{\beta}}{\hbar}Nt_{0}}\nonumber\\
&&\cdot \sum_{\alpha=1}^{2k}
\sum_{\substack{k_1+\cdots+k_c+\frac{d}{2}+\\
s_1+\cdots+s_d=2k\\k_1,\cdots,k_c\geq 1\\s_1,\cdots,s_d\geq 0}}
(-\frac{\sqrt{\beta}}{\hbar}N)^\alpha E^{(k_1, \cdots, k_c|s_1+\frac{1}{2}, \cdots, s_d+\frac{1}{2})}_{(\underbrace{0, \cdots, 0}_\alpha| )}t_{k_1}\cdots t_{k_c}\xi_{s_1+\frac{1}{2}}\cdots \xi_{s_d+\frac{1}{2}},\nonumber
\end{eqnarray}
where we have used the formal expression
of $\tilde W^{l_{2r-1}}\mathbb{\tilde W}^{l_{2r-2}}
\cdots\mathbb{\tilde W}^{l_2}\tilde W^{l_1}$ with degree $2k$
\begin{eqnarray}\label{wep}
\tilde W^{l_{2r-1}}\mathbb{\tilde W}^{l_{2r-2}}
\cdots\mathbb{\tilde W}^{l_2}\tilde W^{l_1}
&=&\sum_{a,c=0}^{2k}\sum_{b,d=0}^{2k+1}
\sum_{\substack{i_1,\cdots,i_a=0\\j_1,\cdots,j_b=0}}^{\infty}
\sum_{\substack{k_1+\cdots+k_c+\\
s_1+\cdots+s_d=\rho\\k_1,\cdots,k_c\geq 1\\s_1,\cdots,s_d\geq 0}}
E^{(k_1, \cdots, k_c|s_1+\frac{1}{2}, \cdots, s_d+\frac{1}{2})}_{(i_1, \cdots, i_a|j_1+\frac{1}{2}, \cdots, j_b+\frac{1}{2})}t_{k_1}\cdots t_{k_c}\nonumber\\
&&\cdot \xi_{s_1+\frac{1}{2}}\cdots \xi_{s_d+\frac{1}{2}}\frac{\partial}{\partial t_{i_1}}\cdots\frac{\partial}{\partial t_{i_a}}\frac{\partial}{\partial \xi_{j_1+\frac{1}{2}}}\cdots\frac{\partial}{\partial \xi_{j_b+\frac{1}{2}}},
\end{eqnarray}
$\rho=\sum_{\mu=1}^a i_{\mu}+\sum_{\nu=1}^b j_{\nu}+2k+\frac{b}{2}-\frac{d}{2}$, the coefficients
$E^{(k_1, \cdots, k_c|s_1+\frac{1}{2}, \cdots, s_d+\frac{1}{2})}_{(i_1, \cdots, i_a|j_1+\frac{1}{2}, \cdots, j_b+\frac{1}{2})}$ are  polynomials
with respect to $i_{\mu}$, $j_{\nu}$, $k_{\bar\mu}$ and $s_{\bar\nu}$,
$\bar\mu=1,\cdots, c$, $\bar\nu=1,\cdots d$.

Similarly, we have
\begin{eqnarray}\label{actt2}
T_2e^{-\frac{\sqrt{\beta}}{\hbar}Nt_{0}}&=&
\frac{1}{k!2^k}\frac{1}{\prod_{i=1}^k(4i-1)}
\prod_{i=1}^{r-1}\prod_{j=1}^{l_{2i}}[4(l_1+\cdots+l_{2i-1}+j)-1]
e^{-\frac{\sqrt{\beta}}{\hbar}Nt_{0}}\nonumber\\
&&\cdot \sum_{\alpha=1}^{2k}
\sum_{\substack{k_1+\cdots+k_c+\frac{d}{2}+\\
s_1+\cdots+s_d=2k\\k_1,\cdots,k_c\geq 1\\s_1,\cdots,s_d\geq 0}}
(-\frac{\sqrt{\beta}}{\hbar}N)^\alpha
F^{(k_1, \cdots, k_c|s_1+\frac{1}{2}, \cdots, s_d+\frac{1}{2})}_{(\underbrace{0, \cdots, 0}_\alpha| )}t_{k_1}\cdots t_{k_c}\xi_{s_1+\frac{1}{2}}\cdots \xi_{s_d+\frac{1}{2}},\nonumber\\
T_3e^{-\frac{\sqrt{\beta}}{\hbar}Nt_{0}}
&=&\frac{1}{k!2^k}\frac{1}{\prod_{i=l_1+1}^k(4i-1)}
\prod_{i=1}^{r-1}\prod_{j=1}^{l_{2i+1}}[4(l_1+\cdots+l_{2i}+j)-1]
e^{-\frac{\sqrt{\beta}}{\hbar}Nt_{0}}\nonumber\\
&&\cdot \sum_{\alpha=1}^{2k}
\sum_{\substack{k_1+\cdots+k_c+\frac{d}{2}+\\
s_1+\cdots+s_d=2k\\k_1,\cdots,k_c\geq 1\\s_1,\cdots,s_d\geq 0}}
(-\frac{\sqrt{\beta}}{\hbar}N)^\alpha
G^{(k_1, \cdots, k_c|s_1+\frac{1}{2}, \cdots, s_d+\frac{1}{2})}_{(\underbrace{0, \cdots, 0}_\alpha| )}t_{k_1}\cdots t_{k_c}\xi_{s_1+\frac{1}{2}}\cdots \xi_{s_d+\frac{1}{2}},\nonumber\\
T_4e^{-\frac{\sqrt{\beta}}{\hbar}Nt_{0}}&=&
\frac{1}{k!2^k}\frac{1}{\prod_{i=1}^k(4i-1)}
\prod_{i=1}^{r}\prod_{j=1}^{l_{2i}}[4(l_1+\cdots+l_{2i-1}+j)-1]
e^{-\frac{\sqrt{\beta}}{\hbar}Nt_{0}}\nonumber\\
&&\cdot \sum_{\alpha=1}^{2k}
\sum_{\substack{k_1+\cdots+k_c+\frac{d}{2}+\\
s_1+\cdots+s_d=2k\\k_1,\cdots,k_c\geq 1\\s_1,\cdots,s_d\geq 0}}
(-\frac{\sqrt{\beta}}{\hbar}N)^\alpha
H^{(k_1, \cdots, k_c|s_1+\frac{1}{2}, \cdots, s_d+\frac{1}{2})}_{(\underbrace{0, \cdots, 0}_\alpha| )}t_{k_1}\cdots t_{k_c}\xi_{s_1+\frac{1}{2}}\cdots \xi_{s_d+\frac{1}{2}},\nonumber
\\
\end{eqnarray}
where $F^{(k_1, \cdots, k_c|s_1+\frac{1}{2}, \cdots, s_d+\frac{1}{2})}_{(\underbrace{0, \cdots, 0}_\alpha| )}$,
$G^{(k_1, \cdots, k_c|s_1+\frac{1}{2}, \cdots, s_d+\frac{1}{2})}_{(\underbrace{0, \cdots, 0}_\alpha| )}$ and
$H^{(k_1, \cdots, k_c|s_1+\frac{1}{2}, \cdots, s_d+\frac{1}{2})}_{(\underbrace{0, \cdots, 0}_\alpha| )}$ are the coefficients of the terms $t_{k_1}\cdots t_{k_c}\xi_{s_1+\frac{1}{2}}\cdots \xi_{s_d+\frac{1}{2}}\frac{\partial^\alpha}{\partial t_0^\alpha}$ in the formal expansions of the corresponding operators
$\mathbb{\tilde W}^{l_{2r-1}}\tilde W^{l_{2r-2}}
\cdots \tilde W^{l_2}\mathbb{\tilde W}^{l_1}$,
$\mathbb{\tilde W}^{l_{2r}}\tilde W^{l_{2r-1}}
\cdots\mathbb{\tilde W}^{l_2}\tilde W^{l_1}$ and
$\tilde W^{l_{2r}}\mathbb{\tilde W}^{l_{2r-1}}
\cdots\tilde W^{l_2}\mathbb{\tilde W}^{l_1}$, respectively.

Using the actions (\ref{T1act}) and (\ref{actt2}),
we may give the coefficients of $t_{k_1}\cdots t_{k_n}\xi_{s_1+\frac{1}{2}}\cdots\xi_{s_m+\frac{1}{2}}$
with $\sum_{\mu=1}^{n}k_{\mu}+\sum_{\nu=1}^ms_{\nu}+\frac{m}{2}=2k$, $k_{\mu}\geq 1$,
$s_{\nu}\geq 0$ in (\ref{operrepn}).
On the other hand, it is easy to give the corresponding coefficients in (\ref{expan}).
Then from the equivalence of these coefficients, we obtain the desired expressions for the correlators
\begin{eqnarray}\label{gcorr}
\tilde C_{k_1,\cdots, k_n}^{s_1+\frac{1}{2},\cdots, s_m+\frac{1}{2}}
&=&\frac{(-1)^{k+\frac{m(m+1)}{2}}n!(-\frac{\hbar}{\sqrt{\beta}})^{n+m}}
{k!2^k\lambda_{(k_1,\cdots,k_n)}}
\sum_{\alpha=1}^{2k}(-\frac{\sqrt{\beta}}{\hbar}N)^\alpha
\tilde P^{(k_1, \cdots, k_n|s_1+\frac{1}{2}, \cdots, s_m+\frac{1}{2})}_{(\underbrace{0, \cdots, 0}_\alpha|\;)},\nonumber\\
\end{eqnarray}
where $k=\frac{1}{2}(\sum_{\mu=1}^{n}k_{\mu}+\sum_{\nu=1}^ms_{\nu}+\frac{m}{2})$,
$k_{\mu}\geq 1$, $s_{\nu}\geq 0$, $\lambda_{(k_1,\cdots ,k_n)}$ is the number of distinct permutations of $(k_1,\cdots ,k_n)$,
\begin{eqnarray}\label{TPG}
&&\tilde P^{(k_1, \cdots, k_n|s_1+\frac{1}{2}, \cdots, s_m+\frac{1}{2})}_{(\underbrace{0, \cdots, 0}_\alpha|\;)}\nonumber\\
&=&\sum_{\sigma_1, \sigma_2}(-1)^{\tau(\sigma_2(s_1+\frac{1}{2}),\cdots,\sigma_2(s_m+\frac{1}{2}))}
\Big[\sum_{r=1}^{\lfloor \frac{k+1}{2} \rfloor}\sum_{\substack{l_1+\cdots+l_{2r-1}=k\\
l_1,\cdots,l_{2r-1}\geq 1}}
\Big(\frac{1}{\prod_{i=l_1+1}^k(4i-1)}\nonumber\\
&&\cdot \prod_{i=1}^{r-1}\prod_{j=1}^{l_{2i+1}}[4(l_1+\cdots+l_{2i}+j)-1]
 E^{(\sigma_1(k_1), \cdots, \sigma_1(k_n)|\sigma_2(s_1+\frac{1}{2}), \cdots, \sigma_2(s_m+\frac{1}{2}))}_{(\underbrace{0, \cdots, 0}_\alpha|\;)}\nonumber\\
&&+\frac{1}{\prod_{i=1}^k(4i-1)}
\prod_{i=1}^{r-1}\prod_{j=1}^{l_{2i}}[4(l_1+\cdots+l_{2i-1}+j)-1]
 F^{(\sigma_1(k_1), \cdots, \sigma_1(k_n)|\sigma_2(s_1+\frac{1}{2}), \cdots, \sigma_2(s_m+\frac{1}{2}))}_{(\underbrace{0, \cdots, 0}_\alpha|\;)}\Big)\nonumber\\
&&+\sum_{r=1}^{\lfloor \frac{k}{2} \rfloor}\sum_{\substack{l_1+\cdots+l_{2r}=k\\
l_1,\cdots,l_{2r}\geq 1}}\Big(\frac{1}{\prod_{i=l_1+1}^k(4i-1)}
 \prod_{i=1}^{r-1}\prod_{j=1}^{l_{2i+1}}[4(l_1+\cdots+l_{2i}+j)-1]\nonumber\\
&&\cdot G^{(\sigma_1(k_1), \cdots, \sigma_1(k_n)|\sigma_2(s_1+\frac{1}{2}), \cdots, \sigma_2(s_m+\frac{1}{2}))}_{(\underbrace{0, \cdots, 0}_\alpha|\;)}
+\frac{1}{\prod_{i=1}^k(4i-1)}
\prod_{i=1}^{r}\prod_{j=1}^{l_{2i}}[4(l_1+\cdots+l_{2i-1}+j)-1]\nonumber\\
&&\cdot H^{(\sigma_1(k_1), \cdots, \sigma_1(k_n)|\sigma_2(s_1+\frac{1}{2}), \cdots, \sigma_2(s_m+\frac{1}{2}))}_{(\underbrace{0, \cdots, 0}_\alpha|\;)}\Big)
\Big].
\end{eqnarray}
Here $\sigma_1$ denotes all the distinct permutations of $(k_1,\cdots ,k_n)$, $\sigma_2$ is all the distinct permutations of  $(s_1+\frac{1}{2},\cdots,s_m+\frac{1}{2})$
and its  inverse number is denoted as $\tau(\sigma_2(s_1+\frac{1}{2}),\cdots ,\sigma_2(s_m+\frac{1}{2}))$.

Let us list the correlators for the case of $k=2$ in (\ref{gcorr}) as follows:
\begin{eqnarray}
&&\tilde C_{4}
=\frac{\hbar^2}{4\beta}N[2\tilde N^2+(1-\beta)\tilde N+4\beta],
\ \ \ \ \ \ \
\tilde C_{3,1}=\frac{3\hbar^2}{2\beta}N\tilde N,\nonumber\\
&&\tilde C_{2,2}=\frac{\hbar^2}{4\beta}N\tilde N(N\tilde N+4),
\ \ \ \ \ \ \ \ \ \ \ \ \ \ \ \ \ \ \
\tilde C_{1,1,1,1}=\frac{3\hbar^2}{\beta}N^2,\nonumber\\
&&\tilde C_{2,1,1}=\frac{\hbar^2}{2\beta}N(N\tilde N+4),
\ \ \ \ \ \ \ \ \ \ \ \ \ \ \ \ \ \ \ \
C_{1,1}^{\frac{1}{2},\frac{3}{2}}
=\frac{\hbar^2}{\beta}N^2,\nonumber\\
&&\tilde C_{2}^{\frac{1}{2},\frac{3}{2}}=\frac{\hbar^2}{2\beta}N(N\tilde N+4),
\ \ \ \ \ \ \ \ \ \ \ \ \ \ \ \ \ \ \ \ \
C_{1}^{\frac{1}{2},\frac{5}{2}}
=\frac{2\hbar^2}{\beta}N,\nonumber\\
&&\tilde C^{\frac{1}{2},\frac{7}{2}}=\frac{\hbar^2}{2\beta}N[3\tilde N+2(1-\beta)],\ \ \ \ \ \ \ \ \ \ \ \ \
\tilde C^{\frac{3}{2},\frac{5}{2}}=-\frac{\hbar^2}{2\beta}N\tilde N.
\end{eqnarray}

\section{Non-Gaussian supereigenvalue model in the Neveu-Schwarz sector}
Let us consider the non-Gaussian supereigenvalue model in the Neveu-Schwarz sector
\begin{eqnarray}\label{nongeven}
Z_{NG}(t,\xi;a,\varepsilon)&=&(\prod_{i=1}^N \int_{-\infty}^{+\infty}
dz_i \int d\theta_i) \Delta_{NS}(z,\theta)^\beta e^{-\frac{\sqrt{\beta}}{\hbar}
\sum_{i=1}^{N}(V_{NS}(z_i,\theta_i)+\tilde V_{NS}(z_i,\theta_i))},
\end{eqnarray}
where $N$ is even,
$
\tilde V_{NS}(z,\theta)=\frac{1}{2p+2}z^{2p+2}+\sum_{k=1}^{2p}\frac{1}{k}a_kz^k
+\sum_{l=0}^{2p}\varepsilon_l z^l\theta,
$
$p\geq 1$, $a_k$ and $\varepsilon_l$ are nonzero bosonic and fermionic coupling constants, respectively.

By applying the changes of integration variables
$(z_i\rightarrow z_i+\epsilon\sum_{n=2p+1}^\infty nt_{n}z_i^{n-2p-1}, \quad \theta_i\rightarrow \theta_i+\frac{1}{2}\epsilon\sum_{n=2p+1}^\infty n(n-2p-1)t_{n}z_i^{n-2p-2}\theta_i
)$
and
$(z_i\rightarrow z_i+\epsilon\sum_{n=2p+1}^\infty (n+\frac{1}{2})
\xi_{n+\frac{1}{2}}z_i^{n-2p-1}\theta_i, \quad \theta_i\rightarrow \theta_i-\epsilon\sum_{n=2p+1}^\infty (n+\frac{1}{2})\xi_{n+\frac{1}{2}}z_i^{n-2p-1}
)$ for the partition function (\ref{nongeven}), we may derive the constraint from the invariance of the integral
\begin{eqnarray}\label{conseven}
(\hat D+\hat W)Z_{NG}(t,\xi;a,\varepsilon)=0,
\end{eqnarray}
where
\begin{equation}\label{hatD}
\hat D=\sum_{n=2p+1}^{\infty}nt_{n}\frac{\partial}{\partial t_n}
+\sum_{n=2p+1}^{\infty}(n+\frac{1}{2})\xi_{n+\frac{1}{2}}\frac{\partial}{\partial \xi_{n+\frac{1}{2}}},
\end{equation}
$\hat W=\hat W_{2p+2}+\sum_{k=2}^{2p+1}\hat W_{k}+
\sum_{l=1}^{2p+1}\hat W_{l+\frac{1}{2}}$, the operators $\hat W_{2p+2}$, $\hat W_{k}$ and $\hat W_{l+\frac{1}{2}}$ are, respectively, given by
\begin{eqnarray}\label{consopeven}
\hat W_{2p+2}&=&\sum_{n=2p+1}^{\infty}nt_{n}\big[
\sum_{k=1}^\infty kt_k\frac{\partial}{\partial t_{n+k-2p-2}}+\frac{\hbar^2}{2}\sum_{k=0}^{n-2p-2}\frac{\partial}{\partial t_k}\frac{\partial}{\partial t_{n-k-2p-2}}\nonumber\\
&&+\sum_{k=0}^\infty(\frac{n-1}{2}+k-p)\xi_{k+\frac{1}{2}}
\frac{\partial}{\partial \xi_{k+n-2p-\frac{3}{2}}}
 +\frac{\hbar^2}{2}\sum_{k=1}^{n-2p-2} k\frac{\partial}{\partial \xi_{n-k-2p-\frac{3}{2}}}\frac{\partial}{\partial \xi_{k-\frac{1}{2}}}\nonumber\\
 &&-\frac{\hbar}{2\sqrt{\beta}}
 (1-\beta)(n-2p-1)\frac{\partial}{\partial t_{n-2p-2}}\big]
+\sum_{n=2p+1}^{\infty}(n+\frac{1}{2})\xi_{n+\frac{1}{2}}
\big[
 \sum_{k=1}^\infty kt_k\frac{\partial}{\partial \xi_{n+k-2p-\frac{3}{2}}}\nonumber\\
 &&+\hbar^2\sum_{k=0}^{n-2p-2}\frac{\partial}{\partial \xi_{k+\frac{1}{2}}}\frac{\partial}{\partial t_{n-k-2p-2}}
 -\frac{\hbar}{\sqrt{\beta}}(1-\beta)(n-2p-1)\frac{\partial}{\partial \xi_{n-2p-\frac{3}{2}}}\nonumber\\
 &&+\sum_{k=0}^\infty\xi_{k+\frac{1}{2}}\frac{\partial}{\partial t_{k+n-2p-1}}
 \big],\nonumber\\
\hat W_{k}&=&a_{2p+2-k}
\sum_{n=2p+1}^{\infty}nt_{n}\frac{\partial}{\partial t_{n-k}}
+a_{2p+2-k}
\sum_{n=2p+1}^{\infty}(n+\frac{1}{2})\xi_{n+\frac{1}{2}}\frac{\partial}{\partial \xi_{n+\frac{1}{2}-k}},\nonumber\\
\hat W_{l+\frac{1}{2}}&=&\varepsilon_{2p+1-l}
\sum_{n=2p+1}^{\infty}n(p-l+\frac{n+1}{2})t_{n}\frac{\partial}
{\partial \xi_{n-\frac{1}{2}-l}}\nonumber\\
&&+\varepsilon_{2p+1-l}
\sum_{n=2p+1}^{\infty}(n+\frac{1}{2})\xi_{n+\frac{1}{2}}\frac{\partial}
{\partial t_{n-l}}.
\end{eqnarray}

The partition function (\ref{nongeven}) is now viewed as a vector in the space $\tilde V$
generated by the basis
\begin{equation}\label{ubasis}
\{t_{k_1}\cdots t_{k_n}\xi_{s_1+\frac{1}{2}}\cdots\xi_{s_m+\frac{1}{2}}|
k_1,\cdots,k_n\geq 1, s_1,\cdots,s_m\geq 0, n, m\in \mathbb{N}\},
\end{equation}
with coefficients on $\mathbb{C}[[t_0, a, \varepsilon]]$. Let us
define the degrees of the coupling constants $a_k$ and $\varepsilon_l$
are $0$.
Since $m\in \mathbb{N}$ in the basis (\ref{ubasis}), the partition function (\ref{nongeven}) is graded from
$0, \frac{1}{2}, 1, \frac{3}{2},\cdots, \infty$. Thus we have the expansion
\begin{eqnarray}\label{decom}
Z_{NG}(t,\xi;a,\varepsilon)
&=&e^{-\frac{\sqrt{\beta}}{\hbar}Nt_{0}}Z_{NG}(a,\varepsilon)
+\sum_{s\in \frac{1}{2}\mathbb{N}_{+}}Z^{(s)}_{NG}(t,\xi;a,\varepsilon)\nonumber\\
&=&e^{-\frac{\sqrt{\beta}}{\hbar}Nt_{0}}\Big[Z_{NG}(a,\varepsilon)
-\frac{\sqrt{\beta}}{\hbar}t_{k_1}C_{k_1}(a,\varepsilon)
-\frac{\sqrt{\beta}}{\hbar}\xi_{s_1+\frac{1}{2}}C^{s_1+\frac{1}{2}}
(a,\varepsilon)\nonumber\\
&&+\frac{1}{2!}
(\frac{\sqrt{\beta}}{\hbar})^2 t_{k_1}t_{k_2}C_{k_1, k_2}(a,\varepsilon)
+(\frac{\sqrt{\beta}}{\hbar})^2
t_{k_1}\xi_{s_1+\frac{1}{2}} C_{k_1}^{s_1+\frac{1}{2}}(a,\varepsilon)\nonumber\\
&&-\frac{1}{2!}(\frac{\sqrt{\beta}}{\hbar})^2
\xi_{s_1+\frac{1}{2}}\xi_{s_2+\frac{1}{2}}C^{s_1+\frac{1}{2}, s_2+\frac{1}{2}}(a,\varepsilon)+\cdots\Big],
\end{eqnarray}
where
\begin{eqnarray}
Z_{NG}(a,\varepsilon)
=(\prod_{i=1}^N \int_{-\infty}^{+\infty} dz_i \int d\theta_i) \Delta_{NS}(z,\theta)^\beta e^{-\frac{\sqrt{\beta}}{\hbar}
\sum_{i=1}^{N}\tilde V_{NS}(z_i,\theta_i)},
\end{eqnarray}
\begin{eqnarray}\label{degeven}
Z^{(s)}_{NG}(t,\xi;a,\varepsilon)
&=&e^{-\frac{\sqrt{\beta}}{\hbar}Nt_0}\Big[\sum_{n=0}^{\infty}
\sum_{m=0}^{\infty}\sum_{\substack{k_1+\cdots +k_n+\\s_1+\cdots+s_m+\frac{m}{2}=s\\k_1,\cdots,k_n\geq 1\\s_1,\cdots,s_m\geq 0}}
\frac{(-1)^{\frac{m(m-1)}{2}}(-\frac{\sqrt{\beta}}{\hbar})^{n+m}}{n!m!}
t_{k_1}\cdots t_{k_n}
\nonumber\\
&&\cdot
\xi_{s_1+\frac{1}{2}}\cdots\xi_{s_m+\frac{1}{2}}
C_{k_1, \cdots, k_n}^{s_1+\frac{1}{2}, \cdots, s_m+\frac{1}{2}}(a,\varepsilon)\Big],
\end{eqnarray}
and the correlators  $C_{k_1, \cdots, k_n}^{s_1+\frac{1}{2}, \cdots, s_m+\frac{1}{2}}(a,\varepsilon)$ are defined by
\begin{eqnarray}\label{decoreven}
C_{k_1, \cdots, k_n}^{s_1+\frac{1}{2}, \cdots, s_m+\frac{1}{2}}(a,\varepsilon)&=&(\prod_{i=1}^N \int_{-\infty}^{+\infty} dz_i \int d\theta_i)
\sum_{\substack{i_1,\cdots, i_n=1\\j_1,\cdots, j_m=1}}^N
z_{i_1}^{k_1}\cdots z_{i_n}^{k_n}
z_{j_1}^{s_1}\theta_{j_1}\cdots
z_{j_m}^{s_m}\theta_{j_m}\nonumber\\
&&\cdot \Delta_{NS}(z,\theta)^\beta e^{-\frac{\sqrt{\beta}}{\hbar}\sum_{i=1}^{N}\tilde V_{NS}(z_i,\theta_i)}.
\end{eqnarray}
It should be noted that the correlators
$C_{k_1, \cdots, k_n}^{s_1+\frac{1}{2}, \cdots, s_m+\frac{1}{2}}(a,\varepsilon)$ with the
fermionic coupling constants are written on the right side of $t_{k_1}\cdots t_{k_n}\xi_{s_1+\frac{1}{2}}\cdots\xi_{s_m+\frac{1}{2}}$ in (\ref{decom})
for the convenience of the following discussions.

Let us further analyze the constraint (\ref{conseven}).
We observe that in (\ref{conseven}) $\hat D=D-\sum_{k=1}^{2p}kt_{k}\frac{\partial}{\partial t_k}
-\sum_{l=0}^{2p}(l+\frac{1}{2})\xi_{l+\frac{1}{2}}\frac{\partial}{\partial \xi_{l+\frac{1}{2}}}$
has degree $0$, $\hat W_{2p+2}$, $\hat W_{k}$ and $\hat W_{l+\frac{1}{2}}$  are operators with degrees
$2p+2$, $k$ and $l+\frac{1}{2}$, respectively. Since $Ker(\hat D+\hat W)$ is no longer one dimensional on $\tilde V$,
the partition function $(\ref{nongeven})$ can not be uniquely determined by the constraint (\ref{conseven}).
On the other hand, there are the additional constraints for the partition function (\ref{nongeven})
\begin{eqnarray}\label{addcons}
&&\frac{\partial}{\partial t_k}Z_{NG}(t,\xi;a,\varepsilon)=k\frac{\partial}{\partial a_k}
Z_{NG}(t,\xi;a,\varepsilon),\quad k=1,\cdots,2p,\nonumber\\
&&\frac{\partial}{\partial \xi_{l+\frac{1}{2}}}Z_{NG}(t,\xi;a,\varepsilon)=
\frac{\partial}{\partial \varepsilon_l}Z_{NG}(t,\xi;a,\varepsilon),\quad l=0,\cdots,2p.
\end{eqnarray}
Substituting (\ref{addcons}) into (\ref{conseven}), we obtain
\begin{equation}\label{ngconsadd}
(D+\mathcal{\hat W})Z_{NG}(t,\xi;a,\varepsilon)=0,
\end{equation}
where $\mathcal{\hat W}=\hat W-\sum_{k=1}^{2p}k^2t_k\frac{\partial}{\partial a_k}
-\sum_{l=0}^{2p}(l+\frac{1}{2})\xi_{l+\frac{1}{2}}
\frac{\partial}{\partial \varepsilon_{l}}$.

Since the operators $D$ and $D+\mathcal{\hat W}$ are invertible on $-\mathcal{\hat W} e^{-\frac{\sqrt{\beta}}{\hbar}Nt_{0}}Z_{NG}(a,\varepsilon)$,
by the constraint (\ref{ngconsadd}), we have
\begin{eqnarray}
\sum_{s\in \frac{1}{2}\mathbb{N}_{+}}^{\infty}Z_{NG}^{(s)}(t,\xi;a,\varepsilon)
=-(D+\mathcal{\hat W})^{-1}\mathcal{\hat W}e^{-\frac{\sqrt{\beta}}{\hbar}Nt_{0}}.
\end{eqnarray}
Hence the partition function (\ref{nongeven}) can be expressed as
\begin{eqnarray}\label{operrepneven}
Z_{NG}(t,\xi;a,\varepsilon)=
\sum_{k=0}^{\infty}(-D^{-1}\mathcal{\hat W})^{k}
e^{-\frac{\sqrt{\beta}}{\hbar}Nt_{0}}
Z_{NG}(a,\varepsilon).
\end{eqnarray}

Similar to the representation for the non-Gaussian Hermitian matrix model presented in Ref. \cite{2007.10354},
we see that $\mathcal{\hat W}$ in (\ref{operrepneven}) is not a homogeneous operator.
Since $\mathcal{\hat W}$ contains the noncommutative operators with degrees ranging from $\frac{1}{2}$ to $2p+2$,
it not only leads to the fact that the partition function (\ref{nongeven}) can not be obtained by acting on elementary functions
with exponents of the operator $\mathcal{\hat W}$, but also makes the handling of
the correlators quite difficult from (\ref{operrepneven}).
We can in principle derive the correlators step by step from (\ref{operrepneven}).

For example, we give some correlators as follows:
\begin{eqnarray}\label{ngcorr1}
&&C^{\frac{1}{2}}(a,\varepsilon)=-\frac{\hbar}{\sqrt{\beta}}
\frac{\partial}{\partial \varepsilon_0}
Z_{NG}(a,\varepsilon), \ \ \ \ \ \ \ \ \ \ \ \ \ \ \ \ \ C_1(a,\varepsilon)=-\frac{\hbar}{\sqrt{\beta}}
\frac{\partial}{\partial a_1}
Z_{NG}(a,\varepsilon),\nonumber\\
&&C_1^{\frac{1}{2}}(a,\varepsilon)=\frac{\hbar^2}{\beta}
\frac{\partial^2}{\partial a_1\partial \varepsilon_0}
Z_{NG}(a,\varepsilon),\ \ \ \ \ \ \ \ \ \ \ \ \ \ \ \
C^{\frac{3}{2}}(a,\varepsilon)=-\frac{\hbar}{\sqrt{\beta}}
\frac{\partial}{\partial \varepsilon_1}
Z_{NG}(a,\varepsilon),\nonumber\\
&&C_2(a,\varepsilon)=-\frac{2\hbar}{\sqrt{\beta}}
\frac{\partial}{\partial a_2}
Z_{NG}(a,\varepsilon),\ \ \ \ \ \ \ \ \ \ \ \ \ \ \ \ \
C_{1,1}(a,\varepsilon)=\frac{\hbar^2}{\beta}
\frac{\partial^2}{\partial a_1^2}
Z_{NG}(a,\varepsilon),\nonumber\\
&&C^{\frac{1}{2},\frac{3}{2}}(a,\varepsilon)=\frac{2\hbar^2}{\beta}
\frac{\partial^2}{\partial \varepsilon_0\partial\varepsilon_1}
Z_{NG}(a,\varepsilon), \ \ \ \ \ \ \ \ \ \ \ \
C_2^{\frac{1}{2}}(a,\varepsilon)
=\frac{2\hbar^2}{\beta}\frac{\partial^2}{\partial a_2
\partial \varepsilon_0}Z_{NG}(a,\varepsilon),
\nonumber\\
&&C_{1,1}^{\frac{1}{2}}(a,\varepsilon)=-(\frac{\hbar}{\sqrt{\beta}})^3
\frac{\partial^3}{\partial a_1^2\partial \varepsilon_0}
Z_{NG}(a,\varepsilon), \ \ \ \ \ \ \
C_{1}^{\frac{3}{2}}(a,\varepsilon)=\frac{\hbar^2}{\beta}
\frac{\partial^2}{\partial a_1\partial \varepsilon_1}
Z_{NG}(a,\varepsilon),\nonumber\\
&&C^{\frac{5}{2}}(a,\varepsilon)=-\frac{\hbar}{\sqrt{\beta}}
\frac{\partial}{\partial \varepsilon_2}
Z_{NG}(a,\varepsilon), \ \ \ \ \ \ \ \ \ \ \ \ \ \ \ \
C_{2,1}(a,\varepsilon)=\frac{4\hbar^2}{\beta}\frac{\partial^2 }{\partial a_1\partial a_2}
Z_{NG}(a,\varepsilon),
\nonumber\\
&&C_{1,1,1}(a,\varepsilon)=-(\frac{\hbar}{\sqrt{\beta}})^3\frac{\partial^3 }{\partial a_1^3}Z_{NG}(a,\varepsilon), \ \ \ \ \ \ \ \ \
C^{\frac{1}{2},\frac{5}{2}}(a,\varepsilon)=\frac{2\hbar^2}{\beta}
\frac{\partial^2}{\partial \varepsilon_0\partial \varepsilon_2}Z_{NG}(a,\varepsilon),
\nonumber\\
&&C_3(a,\varepsilon)=
\left\{
  \begin{array}{ll}
    [-Na_1+\frac{\hbar}{\sqrt{\beta}}(a_2\frac{\partial }{\partial a_1}+\varepsilon_1\frac{\partial}{\partial \varepsilon_0}+2\varepsilon_2\frac{\partial}{\partial \varepsilon_1})]Z_{NG}(a,\varepsilon), & \hbox{$p=1$ ;} \\
\\
    -\frac{3\hbar}{\sqrt{\beta}}
\frac{\partial }{\partial a_3}Z_{NG}(a,\varepsilon), & \hbox{$p\geq 2$ ,}
  \end{array}
\right.\nonumber\\
&&C_{1}^{\frac{1}{2},\frac{3}{2}}(a,\varepsilon)
=-2(\frac{\hbar}{\sqrt{\beta}})^3
\frac{\partial^3 }{\partial a_1\partial \varepsilon_0\partial \varepsilon_1}Z_{NG}(a,\varepsilon).
\end{eqnarray}

\section{Chiral supereigenvalue model in the Neveu-Schwarz sector}
Let us consider the chiral supereigenvalue model in the Neveu-Schwarz sector
\begin{eqnarray}\label{cnspf}
 Z_C&=&\frac{1}{\Lambda_C}(\prod_{i=1}^N \int_{0}^{+\infty} dz_i \int d\theta_i) \Delta_{NS}(z,\theta)^\beta e^{-\frac{\sqrt{\beta}}{\hbar}\sum_{i=1}^{N}(V_{NS}(z_i,\theta_i)+z_i)},
\end{eqnarray}
where $N$ is even, the normalization factor $\Lambda_C$ is given by
\begin{eqnarray}
\Lambda_C=(\prod_{i=1}^N \int_{0}^{+\infty} dz_i \int d\theta_i) \Delta_{NS}(z,\theta)^\beta e^{-\frac{\sqrt{\beta}}{\hbar}
\sum_{i=1}^{N}z_i}.
\end{eqnarray}
The correlators $\bar C_{k_1, \cdots, k_n}^{s_1+\frac{1}{2}, \cdots, s_m+\frac{1}{2}}$ in the chiral supereigenvalue model are defined by
\begin{eqnarray}\label{deccor}
\bar C_{k_1, \cdots, k_n}^{s_1+\frac{1}{2}, \cdots, s_m+\frac{1}{2}}&=&\frac{1}{\Lambda_C}(\prod_{i=1}^N \int_{0}^{+\infty} dz_i \int d\theta_i)
\sum_{\substack{i_1,\cdots, i_n=1\\j_1,\cdots, j_m=1}}^N
z_{i_1}^{k_1}\cdots z_{i_n}^{k_n}
z_{j_1}^{s_1}\theta_{j_1}\cdots
z_{j_m}^{s_m}\theta_{j_m}\nonumber\\
&&\cdot \Delta_{NS}(z,\theta)^\beta e^{-\frac{\sqrt{\beta}}{\hbar}\sum_{i=1}^{N}z_i}.
\end{eqnarray}

Note that (\ref{deccor}) is not convergent when the bosonic variables $z_i$ are integrated from
$-\infty$ to $+\infty$. This is the reason why we consider the chiral case (\ref{cnspf}) instead of the supereigenvalue model (\ref{bspse}).

By the invariance of the partition function (\ref{cnspf}) under
two pairs of the changes of integration variables
$(z_i\rightarrow z_i+\epsilon\sum_{n=1}^\infty nt_{n}z_i^{n}, \quad \theta_i\rightarrow \theta_i+\frac{1}{2}\epsilon\sum_{n=1}^\infty n^2t_{n}z_i^{n-1}\theta_i
)$
and
$(z_i\rightarrow z_i+\epsilon\sum_{n=1}^\infty (n+\frac{1}{2})
\xi_{n+\frac{1}{2}}z_i^n\theta_i, \quad \theta_i\rightarrow \theta_i-\epsilon\sum_{n=1}^\infty (n+\frac{1}{2})\xi_{n+\frac{1}{2}}z_i^n
)$, we obtain
\begin{equation}\label{ccons1}
(\tilde D+\bar W)Z_C=0,
\end{equation}
where
\begin{eqnarray}
\bar{W}&=&\sum_{n,k=1}^{\infty}nkt_nt_k\frac{\partial}{\partial t_{n+k-1}}+
\sum_{n,k=1}^{\infty}n(2k+\frac{n+1}{2})t_n\xi_{k+\frac{1}{2}}
\frac{\partial}{\partial \xi_{n+k-\frac{1}{2}}} \nonumber\\
&&+\frac{1}{2}\sum_{n=1}^{\infty}n^2t_n\xi_{\frac{1}{2}}
\frac{\partial}{\partial \xi_{n-\frac{1}{2}}}
+\frac{\hbar^2}{2}\sum_{n=1}^\infty\sum_{k=0}^{n-1}nt_n
\frac{\partial}{\partial t_{k}}\frac{\partial}{\partial t_{n-k-1}}
 \nonumber\\
&&+\frac{\hbar^2}{2}\sum_{n=2}^\infty\sum_{k=1}^{n-1}nkt_n
\frac{\partial}{\partial \xi_{n-k-\frac{1}{2}}}\frac{\partial}{\partial \xi_{k-\frac{1}{2}}} -\frac{\hbar}{2\sqrt{\beta}}(1-\beta)\sum_{n=1}^\infty n^2t_n
\frac{\partial}{\partial t_{n-1}}\nonumber\\
&&+\hbar^2\sum_{n=1}^\infty
\sum_{k=0}^{n-1}(n+\frac{1}{2})\xi_{n+\frac{1}{2}}
\frac{\partial}{\partial \xi_{k+\frac{1}{2}}}\frac{\partial}{\partial t_{n-k-1}} +\sum_{n=1}^{\infty}\sum_{k=0}^{\infty}(n+\frac{1}{2})\xi_{n+\frac{1}{2}}
\xi_{k+\frac{1}{2}}\frac{\partial}{\partial t_{n+k}} \nonumber\\
&& -\frac{\hbar}{\sqrt{\beta}}(1-\beta)\sum_{n=1}^\infty n(n+\frac{1}{2})\xi_{n+\frac{1}{2}}\frac{\partial}{\partial \xi_{n-\frac{1}{2}}}.
\end{eqnarray}

Some specific types of the correlators can be recursively derived from the constraint (\ref{ccons1}).
Taking the coefficients of $t_1^{l+1}$, $t_1^l\xi_{\frac{1}{2}}\xi_{\frac{3}{2}}$, $t_1^lt_2$ in (\ref{ccons1})
and setting to zero, respectively, we obtain
\begin{equation}\label{cc1}
\bar C_1=\frac{\hbar}{2\sqrt{\beta}}N\tilde N,
\end{equation}
and the recursive relations
\begin{eqnarray}\label{crecur}
\bar C_{\underbrace{1, \cdots, 1}_{l+1}}&=&\frac{\hbar}{2\sqrt{\beta}}
(N\tilde{N}+2l)\bar C_{\underbrace{1, \cdots, 1}_{l}},\label{twc11}\nonumber\\
\bar C^{\frac{1}{2},\frac{3}{2}}_{\underbrace{1,\cdots, 1}_{l}}&=&\frac{\hbar}{\sqrt{\beta}(2l+3)}
[l(2l+5+N\tilde{N})\bar C^{\frac{1}{2},\frac{3}{2}}_{\underbrace{1,\cdots, 1}_{l-1}}+3\bar C_{\underbrace{1,\cdots, 1}_{l+1}}],\label{tx1x3}\nonumber\\
\bar C_{2,\underbrace{1, \cdots, 1}_{l}}&=&\frac{\hbar}{\sqrt{\beta}(l+2)}
[l(l+3+\frac{1}{2}N\tilde{N})\bar C_{2,\underbrace{1, \cdots, 1}_{l-1}}+2\tilde{N}\bar C_{\underbrace{1,\cdots, 1}_{l+1}}]\label{ret21}.
\end{eqnarray}
From (\ref{cc1}) and (\ref{crecur}), we can further derive the correlators
\begin{eqnarray}\label{cspcorr}
\bar C_{\underbrace{1, \cdots, 1}_{l}}&=&(\frac{\hbar}{\sqrt{\beta}})^{l}
\prod_{j=0}^{l-1}(\frac{N\tilde{N}}{2}+j),\nonumber\\
\bar C^{\frac{1}{2},\frac{3}{2}}_{\underbrace{1,\cdots, 1}_{l}}&=&
(\frac{\hbar}{\sqrt{\beta}})^{l+2}\frac{N\tilde N}{2}
\prod_{j=1}^l(\frac{N\tilde{N}}{2}+j+1),\nonumber\\
\bar C_{2,\underbrace{1, \cdots, 1}_{l}}
&=&(\frac{\hbar}{\sqrt{\beta}})^{l+2}\frac{N\tilde{N}^2}{2}
\prod_{j=1}^{l}(\frac{N\tilde{N}}{2}+j+1).
\end{eqnarray}

We observe that the operator $\tilde D$ in (\ref{ccons1}) is the same as in (\ref{ccons}),
and the operator $\bar{W}$ in (\ref{ccons1}) is a homogeneous operator with degree $1$.
Following the similar considerations in the Gaussian supereigenvalue model,
we reach the final expressions for the partition function (\ref{cnspf})
\begin{equation}\label{operrepn1}
Z_C=\sum_{k=0}^{\infty}(-\tilde D^{-1}\bar W)^{k}
e^{-\frac{\sqrt{\beta}}{\hbar}Nt_{0}},
\end{equation}
and the correlators
\begin{eqnarray}\label{ccorr}
\bar C_{k_1,\cdots, k_n}^{s_1+\frac{1}{2},\cdots, s_m+\frac{1}{2}}
&=&\frac{(-1)^{s+\frac{m(m+1)}{2}}n!(-\frac{\hbar}{\sqrt{\beta}})^{n+m}}
{s!\lambda_{(k_1,\cdots,k_n)}}
\sum_{\alpha=1}^{2s}(-\frac{\sqrt{\beta}}{\hbar}N)^\alpha
\bar P^{(k_1, \cdots, k_n|s_1+\frac{1}{2}, \cdots, s_m+\frac{1}{2})}_{(\underbrace{0, \cdots, 0}_\alpha|\;)},\nonumber\\
\end{eqnarray}
where $s=\sum_{\mu=1}^{n}k_{\mu}+\sum_{\nu=1}^ms_{\nu}+\frac{m}{2}$,
$k_{\mu}\geq 1$, $s_{\nu}\geq 0$,
\begin{eqnarray}
&&\bar P^{(k_1, \cdots, k_n|s_1+\frac{1}{2}, \cdots, s_m+\frac{1}{2})}_{(\underbrace{0, \cdots, 0}_\alpha|\;)}\nonumber\\
&=&\sum_{\sigma_1, \sigma_2}(-1)^{\tau(\sigma_2(s_1+\frac{1}{2}),\cdots,\sigma_2(s_m+\frac{1}{2}))}
\Big[\sum_{r=1}^{\lfloor \frac{s+1}{2} \rfloor}\sum_{\substack{l_1+\cdots+l_{2r-1}=s\\
l_1,\cdots,l_{2r-1}\geq 1}}
\Big(\frac{1}{\prod_{i=l_1+1}^s(2i-1)}\nonumber\\
&&\cdot \prod_{i=1}^{r-1}\prod_{j=1}^{l_{2i+1}}[2(l_1+\cdots+l_{2i}+j)-1]
\bar E^{(\sigma_1(k_1), \cdots, \sigma_1(k_n)|\sigma_2(s_1+\frac{1}{2}), \cdots, \sigma_2(s_m+\frac{1}{2}))}_{(\underbrace{0, \cdots, 0}_\alpha|\;)}\nonumber\\
&&+\frac{1}{\prod_{i=1}^s(2i-1)}
\prod_{i=1}^{r-1}\prod_{j=1}^{l_{2i}}[2(l_1+\cdots+l_{2i-1}+j)-1]
\bar F^{(\sigma_1(k_1), \cdots, \sigma_1(k_n)|\sigma_2(s_1+\frac{1}{2}), \cdots, \sigma_2(s_m+\frac{1}{2}))}_{(\underbrace{0, \cdots, 0}_\alpha|\;)}\Big)\nonumber\\
&&+\sum_{r=1}^{\lfloor \frac{s}{2} \rfloor}\sum_{\substack{l_1+\cdots+l_{2r}=s\\
l_1,\cdots,l_{2r}\geq 1}}\Big(\frac{1}{\prod_{i=l_1+1}^s(2i-1)}
\prod_{i=1}^{r-1}\prod_{j=1}^{l_{2i+1}}[2(l_1+\cdots+l_{2i}+j)-1]\nonumber\\
&&\cdot
\bar G^{(\sigma_1(k_1), \cdots, \sigma_1(k_n)|\sigma_2(s_1+\frac{1}{2}), \cdots, \sigma_2(s_m+\frac{1}{2}))}_{(\underbrace{0, \cdots, 0}_\alpha|\;)}
+\frac{1}{\prod_{i=1}^s(2i-1)}
\prod_{i=1}^{r}\prod_{j=1}^{l_{2i}}[2(l_1+\cdots+l_{2i-1}+j)-1]\nonumber\\
&&\cdot \bar H^{(\sigma_1(k_1), \cdots, \sigma_1(k_n)|\sigma_2(s_1+\frac{1}{2}), \cdots, \sigma_2(s_m+\frac{1}{2}))}_{(\underbrace{0, \cdots, 0}_\alpha|\;)}\Big)
\Big].
\end{eqnarray}
Here the four functions $\bar E^{(k_1, \cdots, k_c|s_1+\frac{1}{2}, \cdots, s_d+\frac{1}{2})}_{(\underbrace{0, \cdots, 0}_{\alpha}| )}$,
$\bar F^{(k_1, \cdots, k_c|s_1+\frac{1}{2}, \cdots, s_d+\frac{1}{2})}_{(\underbrace{0, \cdots, 0}_{\alpha}| )}$,
$\bar G^{(k_1, \cdots, k_c|s_1+\frac{1}{2}, \cdots, s_d+\frac{1}{2})}_{(\underbrace{0, \cdots, 0}_{\alpha}| )}$ and
$\bar H^{(k_1, \cdots, k_c|s_1+\frac{1}{2}, \cdots, s_d+\frac{1}{2})}_{(\underbrace{0, \cdots, 0}_{\alpha}| )}$ are, respectively,
the coefficients of $t_{k_1}\cdots t_{k_c}\xi_{s_1+\frac{1}{2}}\cdots \xi_{s_d+\frac{1}{2}}\frac{\partial^{\alpha}}{\partial t_0^{\alpha}}$
in the formal expansions of the corresponding operators
$\bar W^{l_{2r-1}}\mathbb{\bar W}^{l_{2r-2}}
\cdots \mathbb{\bar W}^{l_2}\bar W^{l_1}$,
$\mathbb{\bar W}^{l_{2r-1}}\bar W^{l_{2r-2}}
\cdots \bar W^{l_2}\mathbb{\bar W}^{l_1}$,
$\mathbb{\bar W}^{l_{2r}}\bar W^{l_{2r-1}}
\cdots\mathbb{\bar W}^{l_2}\bar W^{l_1}$ and
$\bar W^{l_{2r}}\mathbb{\bar W}^{l_{2r-1}}
\cdots\bar W^{l_2}\mathbb{\bar W}^{l_1}$,
the operator $\mathbb{\bar W}$ is given by
$\mathbb{\bar W}=\xi_{\frac{1}{2}}\frac{\partial }{\partial \xi_{\frac{1}{2}}}\bar W$.

Let us list the correlators (\ref{ccorr}) with $\sum_{\mu=1}^{n}k_{\mu}+\sum_{\nu=1}^ms_{\nu}+\frac{m}{2}\leq 3$ as follows:
\begin{eqnarray}
&&\bar C_1=\frac{\hbar}{2\sqrt{\beta}}N\tilde N,\qquad\ \ \ \ \ \ \ \ \
\bar C_2=\frac{\hbar^2}{2\beta}N\tilde N^2,\nonumber\\
&&\bar C_{1,1}=\frac{\hbar^2}{4\beta}N\tilde N(N\tilde N+2),\quad
\bar C^{\frac{1}{2},\frac{3}{2}}=\frac{\hbar^2}{2\beta}N\tilde{N},
\nonumber\\
&&\bar C_3=\frac{1}{8}(\frac{\hbar}{\sqrt{\beta}})^3N\tilde N[4\beta+\tilde{N}(1-\beta)+5\tilde N^2],\nonumber\\
&&\bar C_{2,1}=
\frac{1}{4}(\frac{\hbar}{\sqrt{\beta}})^3N\tilde N^2(N\tilde{N}+4),\nonumber\\
&&\bar C_{1,1,1}=
\frac{1}{8}(\frac{\hbar}{\sqrt{\beta}})^3N\tilde N(N\tilde N+2)(N\tilde N+4),\nonumber\\
&&\bar C_{1}^{\frac{1}{2},\frac{3}{2}}=
\frac{1}{4}(\frac{\hbar}{\sqrt{\beta}})^3
N\tilde{N}(N\tilde{N}+4),\nonumber\\
&&\bar C^{\frac{1}{2},\frac{5}{2}}=\frac{1}{2}(\frac{\hbar}{\sqrt{\beta}})^3
N\tilde N(2\tilde N+1-\beta).
\end{eqnarray}

\section{Non-Gaussian chiral supereigenvalue model in the Neveu-Schwarz sector}

The non-Gaussian chiral supereigenvalue model in the Neveu-Schwarz sector is
\begin{eqnarray}\label{nongodd}
Z_{CNG}(t,\xi;a,\varepsilon)&=&(\prod_{i=1}^N \int_{0}^{+\infty} dz_i \int d\theta_i) \Delta_{NS}(z,\theta)^\beta e^{-\frac{\sqrt{\beta}}{\hbar}
\sum_{i=1}^{N}(V_{NS}(z_i,\theta_i)+\bar V_{NS}(z_i,\theta_i))},
\end{eqnarray}
where $N$ is even,
$
\bar V_{NS}(z,\theta)=\frac{1}{2p+1}z^{2p+1}+\sum_{k=1}^{2p}\frac{1}{k}a_kz^k
+\sum_{l=0}^{2p}\varepsilon_l z^l\theta,
$
$p\geq 1$, $a_k$ and $\varepsilon_l$ are nonzero bosonic and fermionic coupling constants, respectively.
The correlators $\hat C_{k_1, \cdots, k_n}^{s_1+\frac{1}{2}, \cdots, s_m+\frac{1}{2}}(a,\varepsilon)$ are defined by
\begin{eqnarray}\label{decorodd}
\hat C_{k_1, \cdots, k_n}^{s_1+\frac{1}{2}, \cdots, s_m+\frac{1}{2}}(a,\varepsilon)
&=&(\prod_{i=1}^N \int_{0}^{+\infty} dz_i \int d\theta_i)
\sum_{\substack{i_1,\cdots, i_n=1\\j_1,\cdots, j_m=1}}^N
z_{i_1}^{k_1}\cdots z_{i_n}^{k_n}
z_{j_1}^{s_1}\theta_{j_1}\cdots
z_{j_m}^{s_m}\theta_{j_m}\nonumber\\
&&\cdot \Delta_{NS}(z,\theta)^\beta e^{-\frac{\sqrt{\beta}}{\hbar}\sum_{i=1}^{N}\bar V_{NS}(z_i,\theta_i)}.
\end{eqnarray}

By applying the changes of integration variables
$(z_i\rightarrow z_i+\epsilon\sum_{n=2p+1}^\infty nt_{n}z_i^{n-2p}, \quad \theta_i\rightarrow \theta_i+\frac{1}{2}\epsilon\sum_{n=2p+1}^\infty n(n-2p)t_{n}z_i^{n-2p-1}\theta_i
)$
and
$(z_i\rightarrow z_i+\epsilon\sum_{n=2p+1}^\infty (n+\frac{1}{2})
\xi_{n+\frac{1}{2}}z_i^{n-2p}\theta_i, \quad \theta_i\rightarrow \theta_i-\epsilon\sum_{n=2p+1}^\infty (n+\frac{1}{2})\xi_{n+\frac{1}{2}}z_i^{n-2p}
)$ for the partition function (\ref{nongodd}), we may derive the constraint from the invariance of the integral
\begin{eqnarray}\label{consodd}
(\hat D+\check W)Z_{CNG}(t,\xi;a,\varepsilon)=0,
\end{eqnarray}
where $\check W=\check W_{2p+1}+\sum_{k=1}^{2p}\check W_k+
\sum_{l=0}^{2p}\check W_{l+\frac{1}{2}}$, the operators $\check W_{2p+1}$,
$\check W_k$ and $\check W_{l+\frac{1}{2}}$ are, respectively, given by
\begin{eqnarray}\label{consopodd}
\check W_{2p+1}&=&\sum_{n=2p+1}^{\infty}nt_{n}\big[
\sum_{k=1}^\infty kt_k\frac{\partial}{\partial t_{n+k-2p-1}}
+\frac{\hbar^2}{2}\sum_{k=0}^{n-2p-1}\frac{\partial}{\partial t_k}\frac{\partial}{\partial t_{n-k-2p-1}}
\nonumber\\
&&+\sum_{k=0}^\infty(\frac{n}{2}+k-p)
\xi_{k+\frac{1}{2}}\frac{\partial}{\partial \xi_{k+n-2p-\frac{1}{2}}}
+\frac{\hbar^2}{2}\sum_{k=1}^{n-2p-1} k \frac{\partial}{\partial \xi_{n-k-2p-\frac{1}{2}}}\frac{\partial}{\partial \xi_{k-\frac{1}{2}}}
 \nonumber\\
 &&-\frac{\hbar}{2\sqrt{\beta}}
 (1-\beta)(n-2p)\frac{\partial}{\partial t_{n-2p-1}}\big]
+\sum_{n=2p+1}^{\infty}
 (n+\frac{1}{2})\xi_{n+\frac{1}{2}}\big[
 \sum_{k=1}^\infty kt_k\frac{\partial}{\partial \xi_{n+k-2p-\frac{1}{2}}}
 \nonumber\\
 &&+\hbar^2\sum_{k=0}^{n-2p-1}\frac{\partial}{\partial \xi_{k+\frac{1}{2}}}\frac{\partial}{\partial t_{n-k-2p-1}}
 -\frac{\hbar}{\sqrt{\beta}}(1-\beta)(n-2p)\frac{\partial}{\partial \xi_{n-2p-\frac{1}{2}}}\nonumber\\
 &&+\sum_{k=0}^\infty\xi_{k+\frac{1}{2}}\frac{\partial}{\partial t_{k+n-2p}}\big],\nonumber\\
\check W_k&=&a_{2p+1-k}
\sum_{n=2p+1}^{\infty}nt_{n}\frac{\partial}{\partial t_{n-k}}
+a_{2p+1-k}
\sum_{n=2p+1}^{\infty}(n+\frac{1}{2})\xi_{n+\frac{1}{2}}\frac{\partial}{\partial \xi_{n+\frac{1}{2}-k}},\nonumber\\
\check W_{l+\frac{1}{2}}&=&\varepsilon_{2p-l}
\sum_{n=2p+1}^{\infty}n(p-l+\frac{n}{2})t_{n}\frac{\partial}
{\partial \xi_{n-\frac{1}{2}-l}}+\varepsilon_{2p-l}
\sum_{n=2p+1}^{\infty}(n+\frac{1}{2})\xi_{n+\frac{1}{2}}\frac{\partial}
{\partial t_{n-l}}.
\end{eqnarray}

In similarity with the case of non-Gaussian supereigenvalue model (\ref{nongeven}),
the partition function (\ref{nongodd}) can not be uniquely determined by the constraint (\ref{consodd}).
We have to consider the additional constraints.
It is noted that the partition function (\ref{nongodd}) also satisfies the relations (\ref{addcons}). Substituting (\ref{addcons}) into (\ref{consodd}), we obtain
\begin{equation}\label{cngaddcon}
(D+\mathcal{\check W})Z_{CNG}(t,\xi;a,\varepsilon)=0,
\end{equation}
where $\mathcal{\check W}=\check W-\sum_{k=1}^{2p}k^2t_k\frac{\partial}{\partial a_k}-\sum_{l=0}^{2p}(l+\frac{1}{2})\xi_{l+\frac{1}{2}}
\frac{\partial}{\partial \varepsilon_{l}}$.

Furthermore, from (\ref{cngaddcon}) we can derive
\begin{eqnarray}\label{operrepnodd}
Z_{CNG}(t,\xi;a,\varepsilon)=
\sum_{k=0}^{\infty}(-D^{-1}\mathcal{\check W})^{k}
e^{-\frac{\sqrt{\beta}}{\hbar}Nt_{0}}
Z_{CNG}(a,\varepsilon),
\end{eqnarray}
where $Z_{CNG}(a,\varepsilon)$ is given by
\begin{eqnarray}
Z_{CNG}(a,\varepsilon)
=(\prod_{i=1}^N \int_{0}^{+\infty} dz_i \int d\theta_i) \Delta_{NS}(z,\theta)^\beta e^{-\frac{\sqrt{\beta}}{\hbar}
\sum_{i=1}^{N}\bar V_{NS}(z_i,\theta_i)}.
\end{eqnarray}

Note that $\mathcal{\check W}$ in (\ref{operrepnodd}) contains noncommutative operators with degrees ranging from $\frac{1}{2}$ to $2p+1$,
it causes that the general expressions of the correlators (\ref{decorodd}) can not be obtained.
In principle, we can calculate the correlators step by step from (\ref{operrepnodd}).

For example, we present the correlators (\ref{decorodd}) with $\sum_{\mu=1}^{n}k_{\mu}+\sum_{\nu=1}^ms_{\nu}+\frac{m}{2}\leq 3$.
They are the same as (\ref{ngcorr1})
by replacing $Z_{NG}(a,\varepsilon)$ by $Z_{CNG}(a,\varepsilon)$,
except for the case of $p=1$ in (\ref{decorodd})
\begin{eqnarray}
\hat C_3(a,\varepsilon)=
    \frac{\hbar}{\sqrt{\beta}}[\frac{N\tilde N}{2}
+\sum_{k=1}^2ka_k\frac{\partial }{\partial a_k}+\sum_{l=0}^{2}(l+\frac{1}{2})\varepsilon_l\frac{\partial}{\partial \varepsilon_l}]Z_{CNG}(a,\varepsilon).
\end{eqnarray}

\section{Conclusions}

The $W$-representations of matrix models may realize the partition functions by
acting on elementary functions with exponents of the given $W$ operators.
They play an important role in the calculations of the correlators.
For the supereigenvalue model in the Ramond sector, one has found its $W$-representation
and derived the compact expressions of the correlators.
In this paper, we have analyzed the Gaussian and chiral supereigenvalue models
in the Neveu-Schwarz sector.
Their partition functions (\ref{gnspf}) and (\ref{cnspf}) can be expressed as the infinite
sums of the homogeneous operators $(-\tilde D^{-1}\tilde W)^k$ and $(-\tilde D^{-1}\bar W)^k$
acting on the function $e^{-\frac{\sqrt{\beta}}{\hbar}Nt_{0}}$, i.e., (\ref{operrepn})
and (\ref{operrepn1}), respectively.
We noted that $\tilde D$ is not the degree operator. It leads to that the usual
$W$-representations of these matrix models can not be provided here.
In spite of this negative result, the notable feature has emerged from the studies of the correlators.
Since the actions of the homogeneous operators $(-\tilde D^{-1}\tilde W)^k$ and $(-\tilde D^{-1}\bar W)^k$
on the function $e^{-\frac{\sqrt{\beta}}{\hbar}Nt_{0}}$ can be evaluated explicitly,
we have derived the compact expressions of the correlators (\ref{gcorr}) and (\ref{ccorr})
from (\ref{operrepn}) and (\ref{operrepn1}), respectively.
Our analysis provides additional insight into these supereigenvalue models.

We have also considered the non-Gaussian (chiral) supereigenvalue models in the Neveu-Schwarz sector.
Unlike the cases of previous Gaussian and chiral supereigenvalue models,
it was noted that the operators $\mathcal{\hat W}$ and $\mathcal{\check W}$ in (\ref{operrepneven})
and (\ref{operrepnodd}) are not the homogeneous operators, which are constituted by the noncommutative
operators with degrees ranging from $\frac{1}{2}$ to $2p+2$ and $2p+1$, respectively.
This makes it quite difficult to derive the compact expressions of the correlators from (\ref{operrepneven})
and (\ref{operrepnodd}).

Finally, it should be mentioned that the partition function for (deformed) Gaussian matrix models
have the explicit character expansions \cite{Alexandrov2014}-\cite{Itoyama2020},
\cite{Itoyama17, 1705.00976, Morozov2018}. The character expansions imply that the expectation values of characters are again characters,
which is considered as the superintegrability property of the models\cite{Mironov2018}.
For the supereigenvalue models, whether there exists the character decomposition formulas deserves further study.

\section *{Acknowledgments}

This work is supported by the National Natural Science Foundation
of China (Nos. 11875194 and 11871350).


\end{document}